\documentclass[prl,twocolumn,superscriptaddress,amsmath,amssymb]{revtex4-2}
\usepackage{graphicx}

\begin{document}

    \title{Optical conductivity and band gap in the double-Weyl candidate SrSi$_{2}$ \\ at ambient pressure}
    \author{L. Z. Maulana}
    \affiliation{1.~Physikalisches Institut, Universit\"at Stuttgart, 70569 Stuttgart, Germany}
    \affiliation{Physics Department, Universitas Jambi, Jambi, Indonesia}
    \author{A. A. Tsirlin}
    \affiliation{Felix-Bloch-Institut f\"ur Festk\"orperphysik, Universit\"at Leipzig, 04103 Leipzig, Germany}
    \author{E. Uykur}
    \affiliation{1.~Physikalisches Institut, Universit\"at Stuttgart, 70569 Stuttgart, Germany}
    \affiliation{Helmholtz-Zentrum Dresden-Rossendorf, Ion Beam Physics and Materials Research, 01328 Dresden, Germany}
    \author{Y. Saito}
    \affiliation{1.~Physikalisches Institut, Universit\"at Stuttgart, 70569 Stuttgart, Germany}
    \author{M. Dressel}
    \affiliation{1.~Physikalisches Institut, Universit\"at Stuttgart, 70569 Stuttgart, Germany}
    \author{M. Imai}
    \affiliation{Research Center for Electronic and Optical Materials, National Institute for Materials Science, 1‑2‑1 Sengen, Tsukuba, Ibaraki 305‑0047, Japan}
    \author{A. V. Pronin}
    \affiliation{1.~Physikalisches Institut, Universit\"at Stuttgart, 70569 Stuttgart, Germany}
    \date{October 5, 2025}

\begin{abstract}

We probe the possible double-Weyl state in cubic SrSi$_{2}$ using optical spectroscopy. The complex optical conductivity was measured in a frequency range from 70~cm$^{-1}$ to 22 000~cm$^{-1}$ at temperatures down to 10 K at ambient pressure. The optical response of SrSi$_{2}$ can be well separated into the intraband (free carriers) and interband contributions. Additionally, four infrared-active phonons are detected. As follows from the optical spectra, the free-carrier density decreases with decreasing temperature, consistent with an activation behaviour. Experimental interband conductivity juxtaposed with ab initio calculations shows that conventional density-functional theory fails to describe the electronic structure of SrSi$_{2}$ in the vicinity of the Fermi level. A semi-local exchange-correlation potential allows a much better agreement with the experiment, resulting in the trivial (gapped) band structure of SrSi$_{2}$. The direct gap estimated from the measurements is approximately 40 meV.

\end{abstract}

\maketitle

\section{Introduction}

The suggestion that one of the strontium-silicide compounds -- cubic SrSi$_{2}$ -- hosts exotic Weyl fermions with a chiral charge of $\pm 2$ \cite{Huang2016} triggered extensive research activities. In this Weyl-semimetal (WSM) state~\cite{Armitage2018}, characterized by simultaneously broken inversion and mirror symmetries, the Weyl nodes of opposite chiralities are situated at different energies, leading to a number of unusual properties, e.g., the quantized circular photogalvanic effect~\cite{deJuan2017}. Furthermore, the WSM state in SrSi$_{2}$ was suggested to be tuned by pressure or doping~\cite{Singh2018}.

Experimental probes of the Fermi surface~\cite{Manna2022} and valence states~\cite{Yao2024}, however, raised doubts about the WSM scenario and suggested that SrSi$_{2}$ may be a narrow-gap semiconductor, as discussed for SrSi$_{2}$ decades ago~\cite{Imai2005}. Still, the size of its band gap has not been determined spectroscopically. Ab initio calculations for SrSi$_{2}$ produced ambiguous results, because details of band dispersions in the vicinity of the Fermi level and even the nature of the ground state (WSM vs. trivial narrow-gap semiconductor) strongly depend on the choice of the exchange-correlation potential~\cite{Manna2022, Imai2025}. Worth noting is the recent prediction of topological phonons in structural enantiomers of SrSi$_{2}$~\cite{Wang2025}.

In addition to the basic-physics interest, SrSi$_{2}$ is also attractive for applications and widely studied as a cost-effective, abundant, and environmentally friendly thermoelectric material~\cite{Hashimoto2007, Kuo2014}. Most recently, composite films and nano-structured samples of SrSi$_{2}$ have been shown to demonstrate enhanced thermometric properties~\cite{Singh2020, Aoyama2023, Ghannam2024}.

In this paper, we report on measurements of frequency-dependent conductivity, $\sigma(\omega) = \sigma_1(\omega) + \textrm{i}\sigma_2(\omega)$, in nominally undoped SrSi$_{2}$ at ambient pressure. We find $\sigma(\omega)$ to demonstrate the behaviour typical for a narrow-gap semiconductor. The measured spectra are decomposed into intra- and interband response and backed up by our first-principles calculations that show a good agreement with the experimental data when a semi-local exchange-correlation potential is used.

Because of large penetration depths, optical (infrared) spectroscopy is a genuine bulk-sensitive technique and is widely used for probing various WSM states~\cite{Pronin2021, Lv2021} as well as for determination of the band gaps in semiconductors~\cite{Yu2010, Dressel2002}. Here, we provide a direct experimental estimate of the band gap in SrSi$_{2}$ that can be used to benchmark ab initio results for this material.

\begin{table*}
\caption{\label{table1}Parameters of the SrSi$_{2}$ samples used in the optical experiments.}

\begin{ruledtabular}
\begin{tabular}{c|c|c|c|c|c|c}
 &{Sr/Si molar ratio}&\multicolumn{4}{c|}{Chemical composition}&{Lattice}\\
 &in the starting material&Sr (wt.~\%)&Si (wt.~\%)&Sr (at.~\%)&Si (at.~\%)&parameter $a$ (\AA)\\ \hline
 Sample 1&1.06:2&61.8(1)&39.0(3)&33.7(1)&66.3(2)&6.537(1)\\
 Sample 2&1.03:2&61.1(3)&39.5(1)&33.1(1)&66.8(1)&6.536(1)\\
\end{tabular}
\end{ruledtabular}
\end{table*}

\section{Materials and Methods}

\begin{figure}[t]
\centering
\includegraphics[width=\columnwidth]{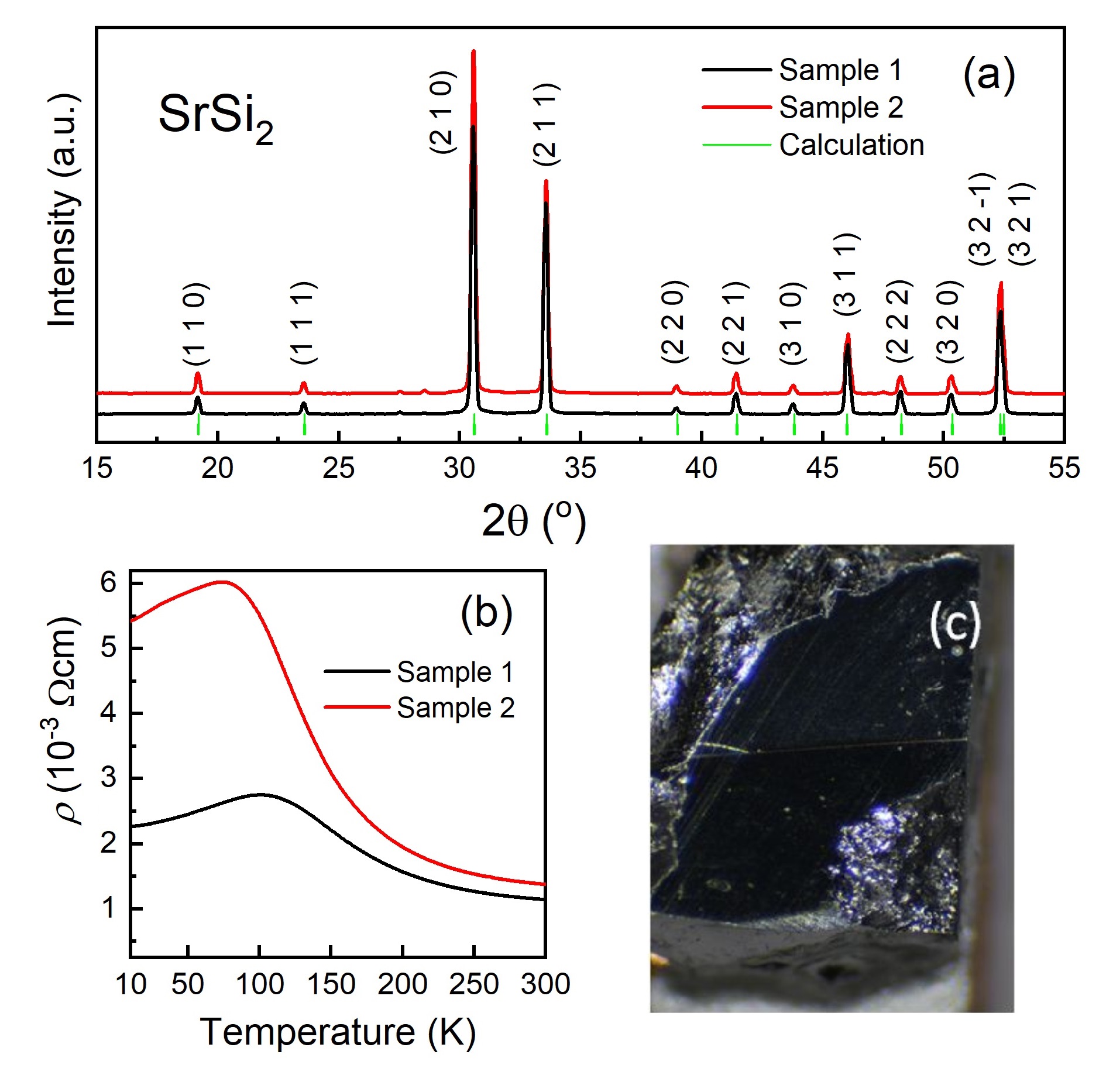}
\caption{(a) Powder XRD patterns of the studied SrSi$_{2}$ samples, (b) their temperature-dependent dc resistivity, and (c) a photograph of one of the samples featuring the surface used for the reflectivity measurements.} \label{rho_xrd}
\end{figure}

Polycrystalline SrSi$_{2}$ samples were prepared by Ar arc melting of a mixture of Sr (Sigma Aldrich, 3N, in ampule) and Si (Furuuchi Chem. Co., 10 N). Table~\ref{table1} lists the molar ratio of Sr to Si in the starting materials and the chemical compositions determined by the inductively coupled plasma optical emission spectroscopy. Figure~\ref{rho_xrd}(a) shows the powder x-ray diffraction (XRD) patterns of the two studied samples. Both patterns are consistent with the pattern simulated for SrSi$_{2}$ using its crystallographic data. The lattice parameters obtained from the experimental XRD patterns, shown in Table~1, are well consistent with that previously reported in Ref.~\onlinecite{Imai2005} (6.5362(3)~{\AA}). To determine the lattice parameter, Si 640c (NIST) was used as an external standard material. The chemical compositions and XRD patterns indicate that the samples consist of single SrSi$_{2}$ phase. Our measurements with a scanning electron microscope show that the diameter of crystallites is in the order of 0.1 mm. The dc electrical resistivity, $\rho(T)$, demonstrates non-monotonic temperature dependence: the resistivity curves possess maxima at intermediate temperatures, as shown in Fig.~\ref{rho_xrd}(b). The temperatures, at which the resistivity has a maximum are approximately 75 and 100 K for the samples 1 and 2, respectively. This temperature dependence of $\rho$ is qualitatively consistent with previous reports~\cite{Imai2005, Imai2025}, although the temperature, at which the resistivity has a maximum, varies somewhat from sample to sample.

Optical reflectivity measurements were conducted on the samples with lateral dimensions of roughly $2 \times 2$ mm$^{2}$ with a
shiny surface (Fig.~\ref{rho_xrd}(c)). The reflectivity $R(\nu)$ was measured at the temperatures from 10 to 295 K over a
frequency range from $\nu = \omega/(2\pi c) = 70$ to $22 000$ cm$^{-1}$ ($\sim$ 8.7 meV -- 2.73 eV) using two Fourier-transform spectrometers. The spectra in the far-infrared ($70 - 700$ cm$^{-1}$) were recorded by a Bruker IFS 113v spectrometer. Here, an in-situ gold evaporation technique was utilized for reference measurements~\cite{Homes1993}. At frequencies above 700 cm$^{-1}$, a Bruker Hyperion microscope attached to a Bruker Vertex 80v spectrometer was used. Freshly evaporated aluminium mirrors served as references in this setup. The complex optical conductivity was obtained from $R(\nu)$ using Kramers-Kronig transformations~\cite{Dressel2002}. High-frequency extrapolations were made utilizing the x-ray atomic scattering functions~\cite{Tanner2015}. At low frequencies, we used the Hagen-Ruben model for the extrapolations.

Density-functional-theory (DFT) band-structure calculations were performed in the Wien2K code~\cite{wien2k,blaha2020} using experimental atomic positions~\cite{evers1978}, whereas the cubic lattice parameter was varied as explained in the following. We utilized two standard DFT functionals, the local density approximation (LDA) by Perdew and Wang~\cite{Perdew1992} and the generalized gradient approximation by Perdew, Burke, and Ernzerhof (PBE)~\cite{Perdew1996}, as well as the semi-local modified Becke-Johnson (mBJ) functional~\cite{tran2009} optimized for reproducing band gaps in semiconductors~\cite{koller2012}. Relativistic calculations were performed on the dense and well-converged $48\times 48\times 48$ $k$-mesh. Optical conductivity was obtained by the internal routine of Wien2K~\cite{draxl2006}.

\section{Results and Discussion}

\begin{figure}[t]
\centering
\includegraphics[width=\columnwidth]{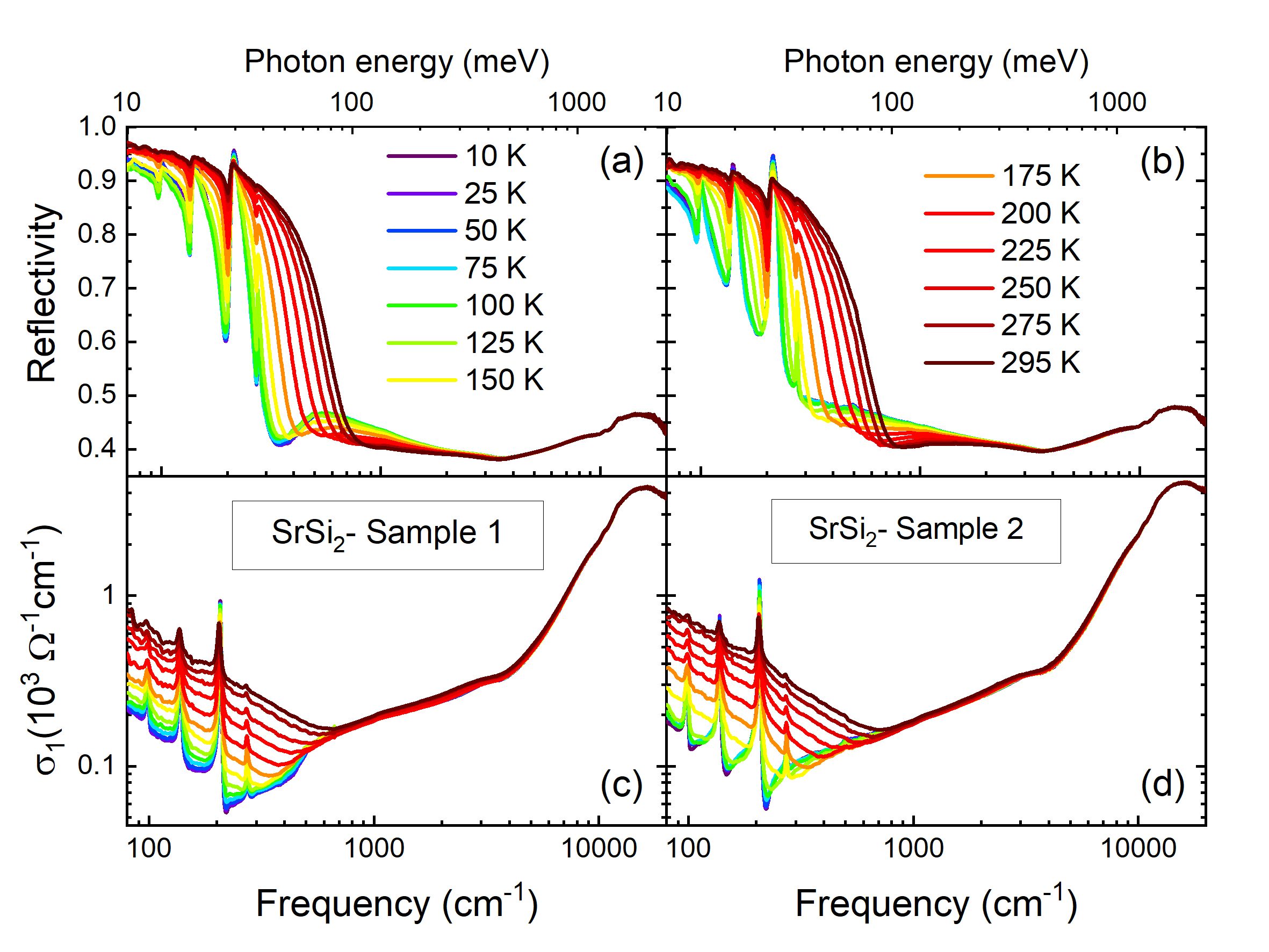}
\caption{(a, b) Reflectivity and (c, d) the real part of optical conductivity for samples 1 (left-hand panels) and 2 (right-hand panels).}
\label{R_sigma}
\end{figure}

Figure~\ref{R_sigma} displays an overview of our optical results. Panels (a) and (b) demonstrate the raw reflectivity for both samples, while in panels (c) and (d) the real part of the optical conductivity is shown. The results for all investigated temperatures are presented. The following conclusions can be immediately drawn by looking at these pictures: (i) both samples demonstrate very similar optical responses; (ii) the material possess free carriers -- the intraband (Drude) response is clearly seen in $R$ and $\sigma_1$; (iii) this response can be clearly separated from the interband absorption setting up at roughly 100 -- 300 cm$^{-1}$, signaling a possible gap in this range; (iv) the interband response is almost temperature independent; (v) the plasma edge of free carriers (the drop in $R(\nu)$) shifts to lower frequencies as $T$ goes down, consistent with a temperature-activated behaviour; (vi) the intensity (the spectral weight) of the Drude mode decreases with decreasing $T$, in agreement with the plasma-edge evolution; (vii) the low-frequency response is affected by four intense phonons, whose positions remain almost perfectly temperature independent, see Table~\ref{table2}. Hereafter, we focus on the electronic response.

\begin{table}[b]
\caption{\label{table2}Eigenfrequencies (in cm$^{-1}$) of the observed transverse infrared-active phonons in SrSi$_{2}$ at 10 and 295 K. The accuracy is within 1~cm$^{-1}$. The values for both samples coincide within this accuracy.}

\begin{ruledtabular}
\begin{tabular}{c|c|c|c|c}
 $T$~(K)&Phonon 1&Phonon 2&Phonon 3&Phonon 4\\ \hline
 10&99&138&208&275\\
 295&99&137&205&271\\
\end{tabular}
\end{ruledtabular}
\end{table}

\begin{figure}[b]
\centering
\includegraphics[width=0.9\columnwidth]{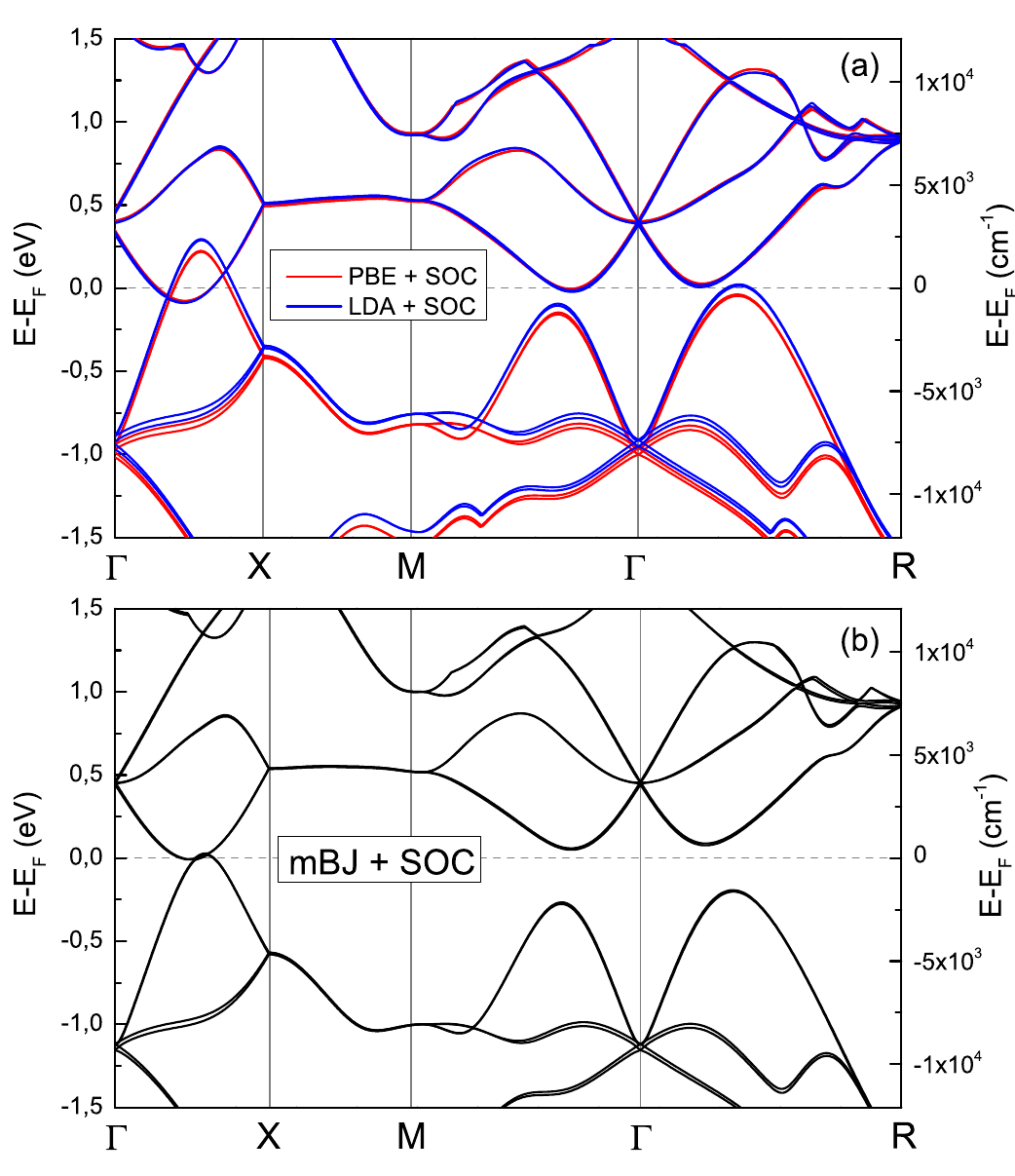}
\caption{Band structures of SrSi$_{2}$ calculated using different approximations for the exchange-correlation potential, as indicated. Experimental lattice parameter of $a=6.535~\textrm{\AA}$ is used~\cite{evers1978}.} \label{bands}
\end{figure}

We first performed a common Drude-Lorentz fit of the spectra~\cite{Dressel2002}. To increase the fit accuracy, we fitted simultaneously three spectra for each temperature (and each sample) -- $R(\nu)$, $\sigma_1(\nu)$, and $\sigma_2(\nu)$. Further, we calculated the SrSi$_{2}$ band structure using three different approaches, as explained above. The obtained band structures are presented in Fig.~\ref{bands}. The important difference between the band structures is that the ones displayed in the upper panel (LDA and PBE) predict band crossings and Weyl nodes on the $\Gamma-X$ line, while the mBJ calculations, shown in the bottom panel, suggest either slightly touching bands or a gapped structure, see the discussion below. Finally, from the band structures, the spectra of interband $\sigma_1(\nu)$ were calculated and compared to our experimental results.

\begin{figure*}[]
\centering
\includegraphics[width=17 cm]{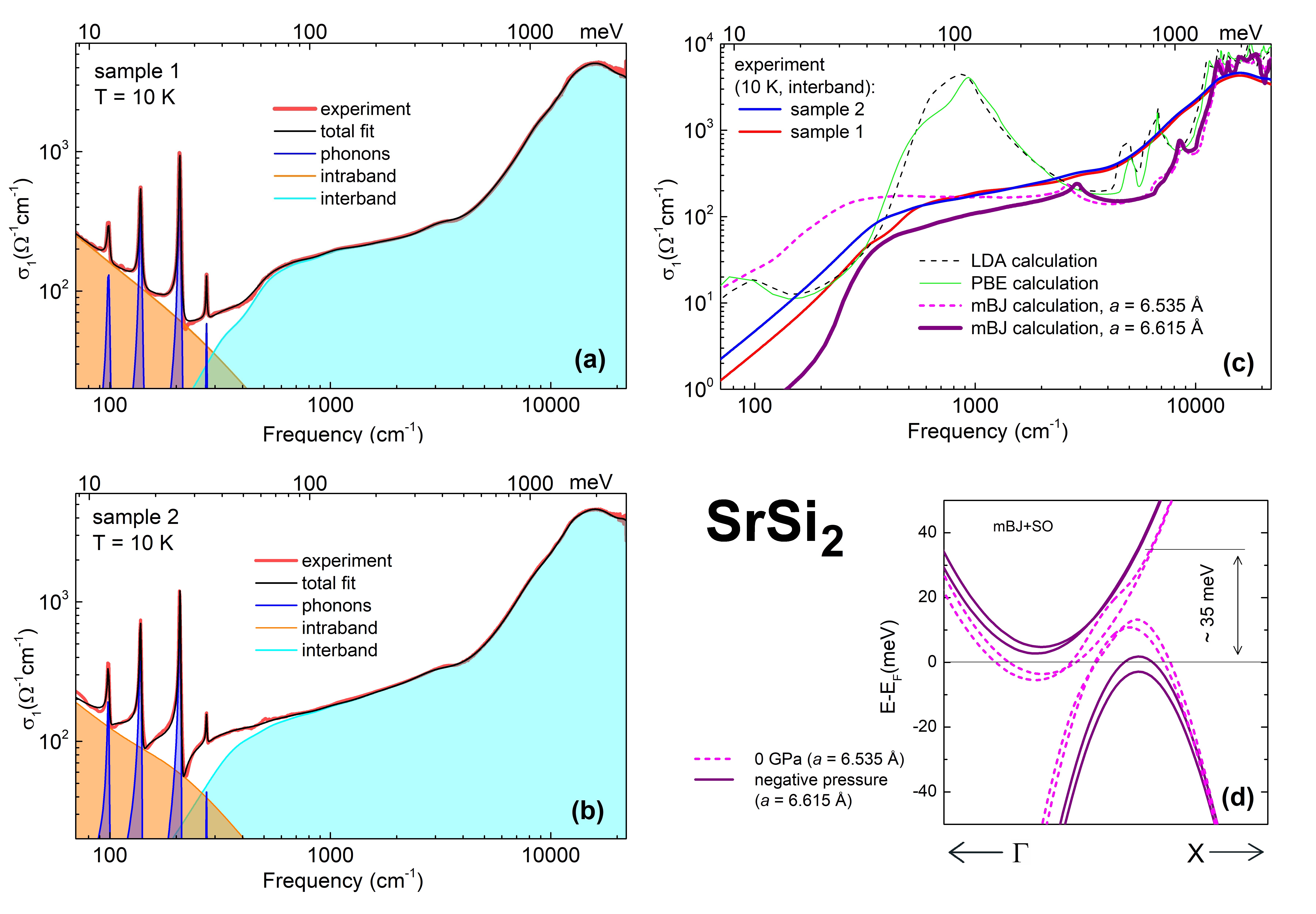}
\caption{(a, b) Experimental conductivity spectra and their Drude-Lorentz fits at 10 K for both samples. (c) Interband part of the experimental conductivity for both samples and the conductivity spectra calculated using three different exchange-correlation functionals (LDA, PBE, mBJ). The increased conductivity above $\sim 0.5$ eV is due to transitions involving multiple (non-parabolic) bands present at higher energies, see Fig.~\ref{bands}. (d) Band structures in the vicinity of the possible band crossing obtained within the mBJ calculations for two lattice parameters $a$.} \label{fits_DFT}
\end{figure*}

In the left-hand panels of Fig.~\ref{fits_DFT} the results of our Drude-Lorentz fits are presented for both samples at 10~K. The experimental spectra can be decomposed into inter- and intraband portions of the electronic response (plus the four sharp phonon peaks mentioned above). The interband contributions are fitted with a number of Lorentzians. We note that in many systems, the intraband conductivity is not represented by a single Drude term, but can effectively be expressed as a sum of one Drude and one finite-frequency Lorentzian component, the latter representing localized carriers. Keeping in mind the polycrystalline structure of our samples, we ascribed the Lorentzian with the lowest eigenfrequency to such a free-carriers mode. Ascribing this Lorentzian to low-energy interband transitions can not be fully excluded. This choice, however, will not change any conclusions drawn below from the comparison of our experiment and the DFT-based conductivity.

Fig.~\ref{fits_DFT}(c) demonstrate this comparison. For both samples, we show the interband portion of the 10-K-data fit and interband conductivities calculated from the three band structures of Fig.~\ref{bands}. It is immediately clear that the LDA and PBE calculations do not reproduce the experiment even on a qualitative level. The huge peaks in $\sigma_1(\nu)$ predicted by these calculations at around 100 meV are not observed experimentally. Instead, the measured spectra are rather flat here, demonstrating only a very modest increase with increasing frequency. Same gentle slope is present in the calculations based on the mBJ band structure, see the magenta and purple curves in Fig.~\ref{fits_DFT}(c). Over the entire frequency window of our experiment, the mBJ curves follow the experimental interband $\sigma_1(\nu)$ quite well. We note that from our experience as well as from the available literature, one cannot expect a perfect match between calculated and measured $\sigma_1(\nu)$, see, e.g., Ref.~\onlinecite{Pronin2021}. We found that the best match between the experimental interband conductivity and the calculations is achieved, if we increase the lattice parameter $a$ by 1.2\%, from 6.535 to 6.615 \AA. Band structures in the vicinity of the possible band crossing are shown in Fig.~\ref{fits_DFT}(d) for these two cases. One can see from Fig.~\ref{fits_DFT}(c) that the slope of $\sigma_1(\nu)$ between 400 and 3000 cm$^{-1}$ is much better reproduced with the expanded lattice parameter than with the experimental one. It tells us that the bands do not cross and gives a very sensitive probe of the band structure in this region. The band structure calculated with the expanded lattice parameter corresponds to the optical gap of 35 meV. Note that this is the direct gap; the indirect gap is as low as 1 meV. The Fermi energy is slightly below the valence band maximum, resulting in a small number of carriers at low temperatures, consistent with the metallic dc resistivity at $T \rightarrow 0$, see Fig.~\ref{rho_xrd}(b). This location of the Fermi energy is consistent with recent photoemission~\cite{Yao2024} and transport~\cite{Imai2025} measurements. No adjustment of the Fermi-level position for the optical-conductivity calculations was made.

By inspecting the conductivity curves shown in Figs.~\ref{R_sigma} and \ref{fits_DFT}, we can provide a direct experimental estimate of the optical (direct) band gap in SrSi$_{2}$: in both samples, the gap can be estimated as roughly 300 cm$^{-1}$ or $\sim 40$ meV. This value is in good agreement with our mBJ gap, but smaller than the recent ARPES result (100 meV) obtained on a single crystal~\cite{Manna2022}. The gap, obtained from the recent transport measurements and the band-structure calculations based on the Heyd–Scuseria–Ernzerhof functional for the samples similar to ours, is in between of these two values (62 meV)~\cite{Imai2025}. We conclude that all recent reports agree on two points: the band structure of SrSi$_{2}$ is gapped and the gap is small, below at least 100 meV.

Earlier DFT studies were inconclusive regarding the presence of the band gap in SrSi$_{2}$ and the size of this possible band gap. Such discrepancies are partially due to the fact that the valence and conduction band may overlap only in the narrow region of the Brillouin zone, along the $\Gamma$-X direction, whereas an indirect gap is always present along $\Gamma$-M and $\Gamma$-R. Additionally, the choice of the DFT functionals affects the positions of the valence and conduction bands. The spurious metallic band structure of SrSi$_{2}$ is obtained with the standard LDA and PBE functionals. Therefore, the use of a non-local or hybrid functional is essential for an accurate treatment of the narrow-gap materials such as SrSi$_{2}$.

\section{Conclusions}

Summarizing, we performed broadband optical-conductivity measurements of cubic SrSi$_{2}$ at ambient pressure and DFT-based band-structure calculations with different exchange-correlation potentials. From the obtained band structures, interband optical conductivity was calculated and compared to the experiment. The comparison shows that the spectra obtained within the mBJ calculations provide reasonable match to the experimental data, while the LDA- and PBE-based spectra do not fit the data at all. Thus, these two approximations can be considered as less relevant for SrSi$_{2}$ and the band structures with crossing electronic bands obtained within these approximations are unlikely to be realized in SrSi$_{2}$ at ambient pressure. The band structure obtained using the mBJ potential does not predict measurable Weyl bands. This provides another evidence for SrSi$_{2}$ being a narrow-gap semiconductor. From the measured spectra and the mBJ-based optical-conductivity calculations, the direct optical gap in SrSi$_{2}$ can be estimated as 35-40 meV at ambient pressure. We finally note that possible WSM state induced by external pressure is discussed in recent theoretical~\cite{Shende2023, Naselli2024} and experimental~\cite{Yao2024} studies of SrSi$_{2}$.

\section{Acknowledgements}

The authors acknowledge fruitful discussions with Weiwu Li and the technical support by Gabriele Untereiner and Di Liu. This work was partly funded by the Deutsche Forschungsgemeinschaft (DFG) via grant No. DR228/51-3 and by the Japan Society for the Promotion of Science (JSPS) via the Grant-in-Aid for Scientific Research program (KAKENHI), grant No. JP22H00268.

\bibliography{SrSi2}

\end{document}